\documentclass[crop]{CSL}

\jname{International Journal of Astrobiology}%

\usepackage{braket}

\begin{document}

\supertitle{Research Article}

\title[Exporting Terrestrial Life]{Exporting Terrestrial Life Out of the Solar System with Gravitational Slingshots of Earthgrazing Bodies}

\author[A. Siraj \& A. Loeb]{Amir Siraj$^{1}$, Abraham Loeb$^{1}$}

\address{\add{1}{Department of Astronomy, Harvard University, 60 Garden Street, Cambridge, MA 02138, USA}}

\corres{\name{Amir Siraj} \email{amir.siraj@cfa.harvard.edu}, \name{Abraham Loeb} \email{aloeb@cfa.harvard.edu}}

\begin{abstract}
Exporting terrestrial life out of the Solar System requires a process that both embeds microbes in boulders and ejects those boulders out of the Solar System. We explore the possibility that Earthgrazing long-period comets and interstellar objects could export life from Earth by collecting microbes from the atmosphere and receiving a gravitational slingshot effect from the Earth. We estimate the total number of exportation events over the lifetime of the Earth to be $\sim 1 - 10$ for long-period comets and $\sim 1 - 50$ for interstellar objects. If life existed above an altitude of 100 km, then the number is dramatically increased up to $\sim 10^5$ exportation events over Earth's lifetime.

\end{abstract}


\selfcitation{Siraj A, Loeb A (2019). Exporting Terrestrial Life Out of the Solar System with Gravitational Slingshots of Earthgrazing Bodies. Submitted to the International Journal of Astrobiology.}

\received{xx xxxx xxxx}

\revised{xx xxxx xxxx}

\accepted{xx xxxx xxxx}

\maketitle

\Fpagebreak

\vfill\pagebreak

\section{Introduction}
Panspermia is the idea that life can propagate from one planet to another \citep{Wesson2010, Wickramasinghe2010}. Impacts on the surface of a planet can launch debris at above the escape speed of the planet, thereby allowing debris spread throughout the planetary system and constituting a plausible mechanism for exchanging life between planets orbiting the same star \citep{Fritz2005, Mileikowsky2000}. However, it is difficult to eject life-bearing material at speeds above the escape speed from a planetary system that is effectively shielded from destructive radiation, presenting a significant challenge for spreading life between stars \citep{Wesson2010}. 

Life in the Earth's atmosphere has been detected up to an altitude of 77 km \citep{Imshenetsky1978}, constituting a reservoir of microbes that objects grazing the atmosphere could draw from. Long-period comets (LPCs) represent a population of bodies that can easily be ejected from the Solar System by gravitational interactions with planets due to their low gravitational binding energies and planet-crossing orbits. This makes them ideal, in principle, for both picking up life from Earth and exporting it out of from the Solar System.

In addition, the high speed and abundance of interstellar objects (ISOs) make them, in addition to LPCs, potential exporters of life from Earth to exoplanetary systems. 1I/`Oumuamua \citep{Meech2017, Micheli2018} was the first interstellar object detected in the Solar System, CNEOS 2014-01-08 \citep{Siraj2019a} was tentatively the first interstellar meteor discovered larger than dust, and 2I/Borisov \citep{Guzik2019} was the first confirmed interstellar comet. \cite{Ginsburg2018} and \cite{Lingam2018} demonstrate dynamically that ejected objects can be gravitationally captured by other star systems.

In this paper, we study whether it is possible for ISOs and LPCs to have exported life from Earth's atmosphere out of the solar system. First, we analyze the gravitational slingshot effect of the Earth during such encounters. We then evaluate the effects of atmospheric drag on the transporting body's size and minimum encounter altitude. Next, we discuss the collection of microbial life during the transporting body's passage through the atmosphere. We then estimate the total number of exportation events since the dawn of life on Earth. Finally, we summarize our main conclusions.

\section{Gravitational Slingshots with Earth}

To gain an understanding of the approximate change in energy that a transporting body receives through a random gravitational interaction with the Earth, we developed a Python code that randomly initializes and integrates the motions of particles from their points of closest approach to Earth in the past or future, computing the total change in energy over the interaction. The Python code created for this work used the open-source N-body integator software \texttt{REBOUND}\footnote{https://rebound.readthedocs.io/en/latest/} to trace the motions of particles under the gravitational influence of the Earth-Sun system \citep{Rein2012}.

We initialize the simulation with the Sun, Earth, and a volume of test particles surrounding Earth at 80 km from the Earth's surface, with near-zero gravitational binding energies from the Sun as appropriate for LPCs.\footnote{ISOs are not included in the simulation, as they are assumed to have binding energies significantly below zero.} The Sun and Earth define the ecliptic plane. For each test particle, we randomly pick an angle within the ecliptic plane between 0 and $2\pi$, as well as a zenith angle between 0 and $\pi$. Using these two angles, we set each particle's position vector relative to Earth. 

To ensure that each particle is at its distance of closest approach, we require the velocity vector to lie in the plane perpendicular to the position vector relative to Earth. For each particle, we pick a random angle between 0 and $2\pi$ to determine in which direction the velocity vector points within this plane. Using the angle within the plane perpendicular to the position vector, we construct each particle's velocity vector. At this point, we have fully initialized the 6D coordinates of each particle in both position and velocity.

\begin{figure}
  \centering
  \includegraphics[width=0.9\linewidth]{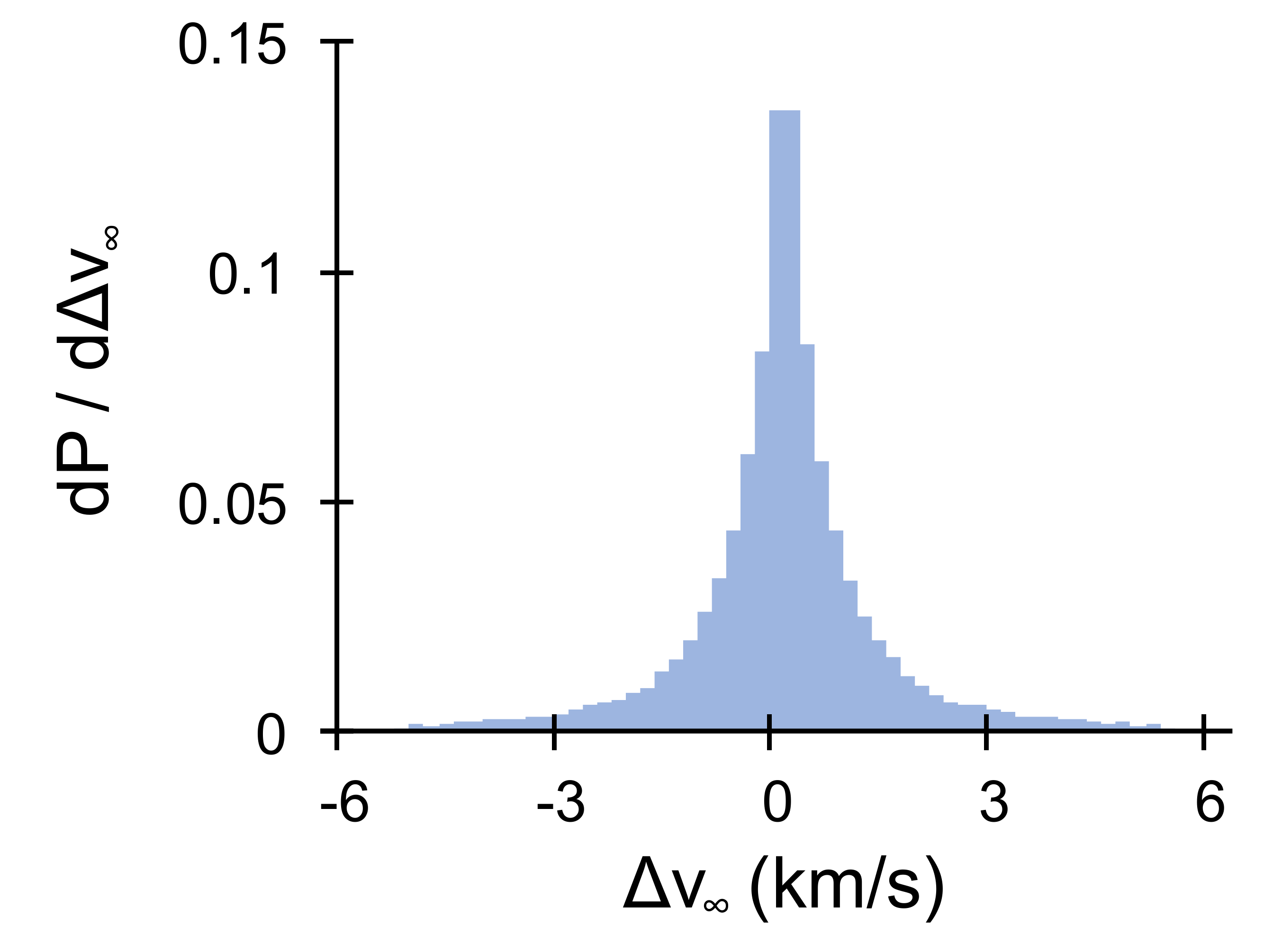}
    \caption{Distribution of the gravitational $\Delta v_{\infty}$ for random encounter geometries between Earth and LPCs marginally bound to the Sun near the Earth's surface.}
    \label{fig:fig1}
\end{figure}

In the first stage of the simulation, we integrate all of the particles backward in time. We use the \texttt{IAS15} integrator in \texttt{REBOUND} to trace each particle from $t = 0$ to an earlier time $-t_i$ \citep{Rein2014}, where $t_i$ is an amount of time to sample either side of the closest approach to Earth. The only constraint on $t_i$ is that it is a time interval at and above which the results of the simulation do not change, on the order of a few times the encounter period; in this case, $t_i$ is on the order of a few days. We record the change in the speed at infinity for the incoming segment of the particle's trajectory, $\Delta v_{\infty, \mathrm{\; in}}$. In the second stage of the simulation, we integrate the particles with unbound initial conditions forward in time. We use \texttt{IAS15} to integrate each particle from $t = 0$ to $t_i$. We record the change in the speed at infinity for the outgoing segment of the particle's trajectory, $\Delta v_{\infty, \mathrm{\; out}}$, and add it to the the incoming and outgoing changes to find the change in speed at infinity for the entire encounter, $\Delta v_{\infty}$. 

We ran our Python code for $10^5$ particles, and the resulting distribution of $\Delta v_{\infty}$ is shown in Figure~\ref{fig:fig1}. Half of encounters result a positive change in energy, as expected from symmetry to time-reversal, and $95 \%$ of such encounters result in $\Delta v_{\infty} \leq 3 \mathrm{\; km \; s^{-1}}$. This corresponds to objects with perihelion distances $\gtrsim 200 \; \mathrm{AU}$, or LPCs.

\section{Atmospheric Drag}

As the transporting object grazes the atmosphere, it encounters atmospheric drag, giving rise to constraints on its minimum size and minimum encounter altitude for it to escape.

Because the change in energy due to the gravitational slingshot is small relative to the initial kinetic energy, we approximate the transporting object's path as linear, summarized by the following expression for altitude,

\begin{equation}
    z(x) = R_{\oplus} + z_{min} - \sqrt{R_{\oplus}^2 - x^2} \; \;,
    \label{eq:1}
\end{equation}
where $R_{\oplus}$ is the radius of the Earth, $z_{min}$ is the minimum altitude of the encounter, and $x$ is a distance parameter that fulfills $z(0) = z_{min}$ and $\mathrm{d}x = v \;\mathrm{d}t$, where $v$ is the instantaneous speed. 

The density of the atmosphere as a function of altitude is $\rho(z) = e^{-z/\mathrm{8 \; km}} \; \mathrm{kg \; m^{-3}}$, and the density of the transporting body is taken to be that of a typical cometary nucleus, $\rho_{\mathrm{tb}} = 600 \; \mathrm{kg \; m^{-3}}$. The acceleration of the transporting body is given by the drag equation \citep{Collins2010},

\begin{equation}
    \frac{\mathrm{d}v}{\mathrm{d}t} = -\frac{3 \rho(z) C_D}{4 \rho_{\mathrm{tb}} L_{\mathrm{tb}}} v^2 \; \; ,
    \label{eq:2}
\end{equation}
where $C_D$ is the drag coefficient, set to the typical value of 2, and $L_{\mathrm{tb}}$ is the length of transporting body.

We use the expression given in \cite{Collins2010} to estimate the yield strength $Y_{\mathrm{tb}}$ of the transporting body to be $\sim 4300 \; \mathrm{Pa}$. The altitude at which the body begins to break up, $z_{\star}$, is given by the solution to the transcendental equation,

\begin{equation}
    Y_{\mathrm{tb}} = \rho(z_{\star}) \; v^2(z_{\star}) \; \; ,
    \label{eq:3}
\end{equation}
which yields, $z_{\star} \simeq \mathrm{100 \; km}$.

\subsection{Expansion \& Slow-down}
At $z < 100 \mathrm{\; km}$, the transporting body expands according to the equation \citep{Collins2010},

\begin{equation}
    \frac{\mathrm{d^2}L_{\mathrm{tb}}}{\mathrm{d}t^2} = \frac{\rho(z) C_D}{\rho_{\mathrm{tb}} L_{\mathrm{tb}}} v^2 \; \; .
    \label{eq:4}
\end{equation}
We note that equations (\ref{eq:2}) -- (\ref{eq:4}) represent the "pancake" or "liquid drop" model, and become less accurate for km-sized impactors and larger, so our results regarding km-sized impactors should be regarded with this caveat \citep{Register2017}.

\begin{figure}
  \centering
  \includegraphics[width=0.9\linewidth]{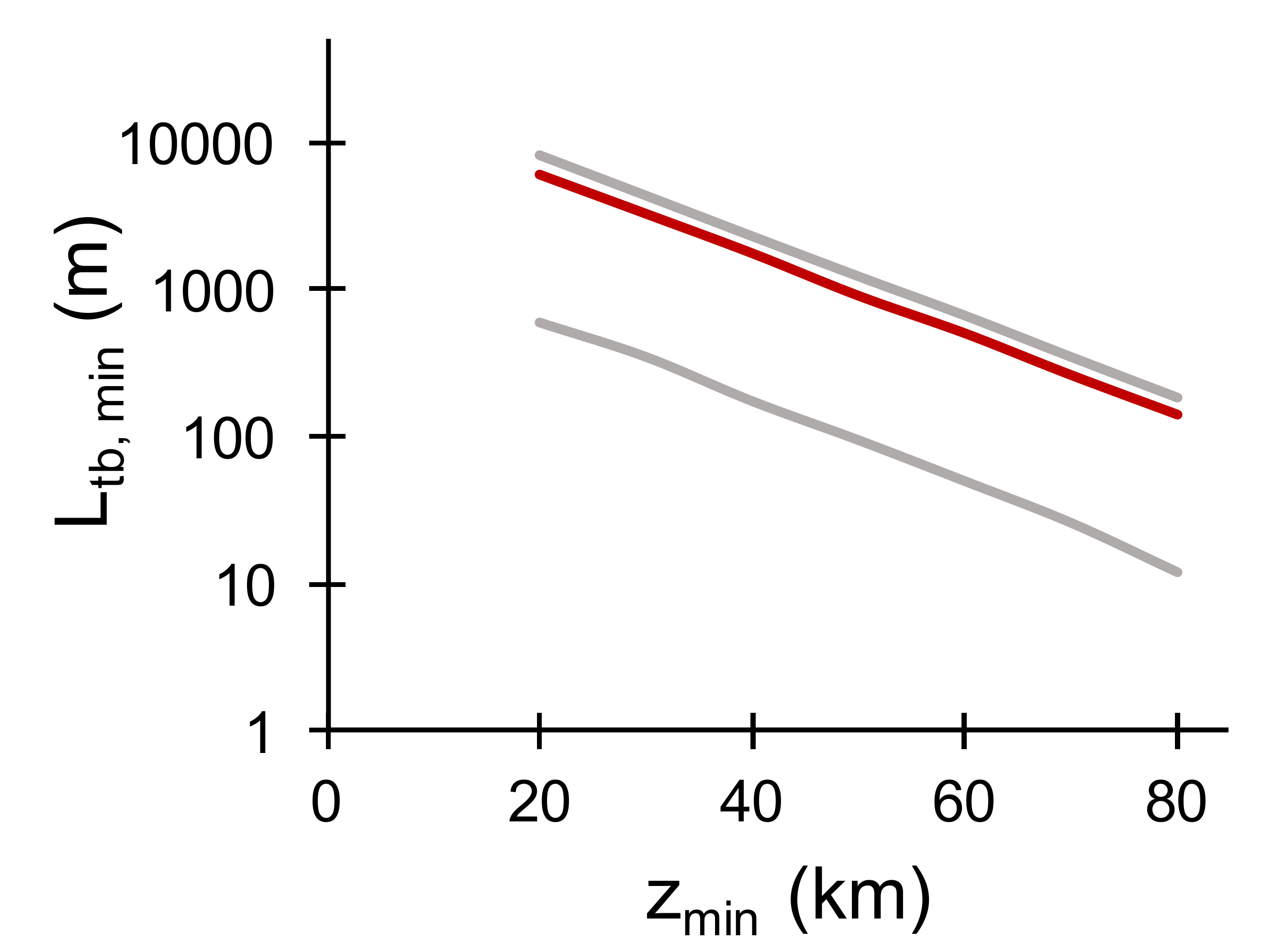}
    \caption{Minimum diameter of transporting body to guarantee total expansion of $< 10\%$ as a function of minimum encounter altitude. The red line indicates the result for an encounter velocity of $42 \mathrm{\; km \; s^{-1}}$, and the gray lines represent the upper and lower bounds, corresponding to the maximum and minimum encounter velocities of $72 \mathrm{\; km \; s^{-1}}$ and $12 \mathrm{\; km \; s^{-1}}$, respectively.}
    \label{fig:fig2}
\end{figure}

We developed Python code that integrates along the path of the transporting body, using equations~(\ref{eq:1}), (\ref{eq:2}), and (\ref{eq:4}) to continuously update $v$ and $L_{\mathrm{tb}}$, thereby deriving the total expansion of the object and change in speed as a function of minimum encounter altitude and size. The minimum size to guarantee a negligible expansion of $\lesssim 10\%$ as a function of altitude is shown in Figure~\ref{fig:fig2} for the range of minimum encounter altitudes 20 km -- 80 km. The lower bound of 20 km was defined by the result that the minimum size was of order the altitude for lower altitudes. The upper bound, 80 km, was chosen because it is the highest altitude at which life has been detected as of yet \citep{Imshenetsky1978}. It is important to note that the \cite{Imshenetsky1978} result should be treated with caution due to the lack of detail on how the system was sterilized and due to the fact that non-biological particles may resemble bacteria, and thus be mistaken for the latter \citep{Wainwright2004, Smith2013}.

The change in speed for all successful encounters is $\lesssim 10^{-5} \; \mathrm{m \; s^{-1}}$, which is reasonable considering the fact that significant slow-down will lead to runaway expansion. Such a small change in speed does not cause significant heating of the body.

\section{Collection of Microbial Life}

While the abundance of  microbes in the upper atmosphere is poorly constrained \citep{Burrows2009}, we use the \cite{Imshenetsky1978} detections to come up with an order-of-magnitude estimate. \cite{Imshenetsky1978} reported 31 colonies of microorganisms collected over 4 sounding rocket launches, each with $\sim 30 \mathrm{\; s}$ at the altitude range 48 -- 77 km, traveling at $\sim 0.75 \mathrm{\; km \; s^{-1}}$, with a detector of size $\sim 5 \times 10^{-3} \mathrm{\; m^2}$. The total detection volume for all 4 flights was then $\sim 450 \;\mathrm{m^3}$, giving an average colony number density of $\sim 0.1 \;\mathrm{m^{-3}}$ at the altitude range 48 -- 77 km.

The total number of collected colonies is then estimated by the equation,

\begin{equation}
    \sim 10^9 \; \mathrm{colonies} \; \left( \frac{L_{\mathrm{tb}}}{400 \; \mathrm{m}} \right)^2 \left( \frac{v}{42 \; \mathrm{km \; s^{-1}}} \right) \left( \frac{\tau_{min}}{2 \; \mathrm{s}} \right) \; \; ,
\end{equation}
where $\tau_{min}$ is approximately the amount of time spent at $z_{min}$. This suggests a large number of collected colonies for typical values. 

The collected microbes will experience accelerations of order $10^5 \; \mathrm{g}$ if they are accelerated over a distance of $\sim 100 \; \mathrm{m}$. For microbes including \textit{Bacillus subtilis}, \textit{Deinococcus radiodurans}, \textit{E. coli}, and \textit{P. denitrificans}, a large proportion would survive accelerations of $4 - 5 \times 10^5 \; \mathrm{g}$, which are of interest for planetary impacts \citep{Mastrapa2001, Deguchi2011}. We therefore assume that acceleration is not an important lethal factor for microbes picked up by the transporting body.

In addition, comet nuclei are porous, making it likely that some incident microbes become embedded several meters below the surface, providing protection from harmful radiation in interstellar space \citep{Wesson2010}.

\section{Number of Exportation Events}

\cite{Francis2005} estimate an LPC flux of $F_{LPC} \sim 11 \; \mathrm{LPCs \; yr^{-1}}$ with $q < \mathrm{4 \; AU}$ and $H \lesssim \mathrm{11}$. \cite{Francis2005} estimate $H \sim \mathrm{11}$ to correspond to a cometary diameter of $L_{H \sim 11} \sim 1 - 2.4 \mathrm{\; km}$ \citep{Bailey1988, Weissman1990}, but \cite{Fernandez2012} estimate a cometary diameter of $L_{H \sim 11} \sim 0.6 \mathrm{\; km}$. \cite{Weissman2007} calculate the impact probability of LPCs with the Earth to be $P_{\oplus} \simeq 2.2 \times 10^{-9}$ per comet per perihelion. While the cumulative size distribution of sub-km LPCs in uncertain, \cite{Vokrouhlicky2019} estimate a cumulative power-law distribution of sizes with an index of $\simeq -1.5$.

We estimate the total number of exportation events caused by Solar System bodies, $N_{S}$, over the age of the Earth, $T_{\oplus}$, given the cross-sectional area of the Earth, $A_{\oplus}$, and occurring between altitudes $z_1$ and $z_2$, to be,

\begin{equation}
    N_{S} \sim F \; P_{\oplus}  T_{\oplus}  A_{\oplus}^{-1} \int^{R_{\oplus} + z_2}_{R_{\oplus} + z_1} \left( \frac{L_{tb, \; min}}{L_{H \sim 11}} \right)^{-1.5} 2 \pi r \; \mathrm{d}r \; ,
\end{equation}
where the value of $L_{tb, \; min}$ is computed as a function of altitude, $z$, with the encounter speed assumed to be $42 \mathrm{\; km \; s^{-1}}$. For $z_1 = 20 \mathrm{\; km}$ and $z_2 = 80 \mathrm{\; km}$, $N_{S} \sim 1 - 10$.

We also consider ISOs, such as `Oumuamua \citep{Meech2017, Micheli2018}. In the $\sim 100 \mathrm{\; m}$ size regime, the power law exponent for the cumulative size distribution is estimated to be $\simeq -3$ \citep{Siraj2019b}. We can then express the total number of exportation events caused by ISOs, $N_{I}$, in terms of the size of `Oumuamua, $L_O$, and the timescale over which an `Oumuamua-size object is expected to collide with the Earth, $t_O^{-1}$,

\begin{equation}
    N_{I} \sim  T_{\oplus}  A_{\oplus}^{-1} t_{O}^{-1} \int^{R_{\oplus} + z_2}_{R_{\oplus} + z_1} \left( \frac{L_{tb, \; min}}{L_O} \right)^{-3} 2 \pi r \; \mathrm{d}r \; ,
\end{equation}
assuming that the composition of such objects are similar to LPCs and that the gravitational deflection by the Earth is small relative to the object's incoming energy.

`Oumuamua's diameter is estimated to be $L_O \simeq 100 - 440 \mathrm{\;m}$, based on Spitzer Space Telescope constraints on its infrared emission given its temperature \citep{Trilling2018}. The implied timescale for collisions of `Oumuamua-size ISOs with the Earth is $t_O \sim 3 \times 10^7 \mathrm{\; yr}$ \citep{Do2018}. As a result, for $z_1 = 20 \mathrm{\; km}$ and $z_2 = 80 \mathrm{\; km}$, $N_{I} \sim 1 - 50$.

\subsection{Possibility of Exportation Events Above 100 km}

Life has not yet been detected above an altitude of 80 km. One obstacle to life in the thermosphere is that the temperature exceeds 400 K, a temperature at which no functional microbes have been confirmed to survive on Earth. However, if we extrapolate the \cite{Imshenetsky1978} results to $z \gtrsim 100 \mathrm{km}$ by assuming that turbulent mixing makes the exponential scale height for microbial life equal to that of air, $\simeq 8 \mathrm{\; km}$, breakup will not occur for icy objects, allowing for smaller and therefore more abundant transporting bodies. If we require that the total number of collected colonies $N_{col} \gtrsim 10^3$, we can derive an expression for $L_{tb, \; min}^{\star}$ as a function of $z$,

\begin{equation}
    L_{tb, \; min}^{\star} \simeq \frac{2}{5} \sqrt{\frac{e^{(z - 63 \mathrm{\; km})/8}}{(v / 42 \mathrm{\; km \; s^{-1}})(t / 2 \mathrm{\; s})}} \; \; .
\end{equation}
The total number of $z \gtrsim 100 \mathrm{\; km}$ exportation events for Solar System bodies and ISOs, respectively, are,

\begin{equation}
    \begin{aligned}
    & N_{S}^{\star} \sim N_S\{z_1 = 100 \mathrm{\; km}, z_2 = \infty, L_{tb, \; min}^{\star}\} \; , \\
    & N_{I}^{\star} \sim N_I\{z_1 = 100 \mathrm{\; km}, z_2 = \infty, L_{tb, \; min}^{\star}\} \; .
    \end{aligned}
\end{equation}
We find $N_{S}^{\star} \sim 3 \times 10^2 - 2 \times 10^3$ and $N_{I}^{\star} \sim 10^3 - 10^5$.

\section{Conclusions}
In this paper we evaluated the possibility of LPCs and ISOs exporting life from Earth's atmosphere out of the Solar System. We estimate the total number of exportation events over the lifetime of the Earth to be $\sim 1 - 10$ for LPCs and $\sim 1 - 50$ for ISOs. If life existed above an altitude of 100 km, we find that up to $\sim 10^5$ exportation events could have occured over Earth's lifetime.

An important comparison to make is to the conventional mode of panspermia involving impacts and subsequent ejecta. \cite{Belbruno2012} find that $10^{14} - 3 \times 10^{16}$ objects with mass $> 10 \mathrm{\; kg}$ were transferred from the Sun to its nearest neighbors in the birth cluster. Assuming a density of $600 \; \mathrm{kg \; m^{-3}}$ and a cumulative size distribution exponent of $-1.5$ in terms of size, we find that $\sim 10^8 - 10^{10}$ km-sized objects were transferred. If $\gtrsim 10^{-6}$ of such objects had viable microbes \citep{Belbruno2012}, this would yield $> 10^2 - 10^4$ objects, which is higher than the number of LPCs and ISOs capable of transferring life, given the cutoff height of 80 km.

The atmospheric scale height, $h \sim \braket{v^2}/g$, where $\braket{v^2}$ includes the sound and turbulence speeds summed in quadrature, and $g$ is the acceleration due to the Earth's gravity. If the atmosphere temperature was higher due to volcanic activity early on, or if turbulence was stronger, then $h$ could have been larger, making the prospect of interstellar panspermia even more realistic.

Improved measurements of the size distribution of LPCs and ISOs would allow for more precise estimates. In addition, more research into the abundance of characteristics of microbes in the upper atmosphere, as well as into impacts at tens of $\mathrm{km \;s^{-1}}$ with such microbes, is crucial for evaluating the merit of the panspermia hypothesis. In particular, the discovery of life above 100 km in the atmosphere would be a very encouraging sign for the feasibility of interstellar panspermia.

\ack[Acknowledgement]{This work was supported in part by a grant from the Breakthrough Prize Foundation.}

\end{document}